# A Novel Graphene-Based Circulator with Multi-layer Triangular Post for THz Region

Mohammad Bagher Heydari [1], Mohammad Hashem Vadjed Samiei [1,*]

[1,*] School of Electrical Engineering, Iran University of Science and Technology (IUST), Tehran, Iran

[*]Corresponding author: mh_samiei@iust.ac.ir

**Abstract:** This article proposes a novel three-port circulator with a triangular graphene-based post for the THz region. This new circulator is formed by three $120^0$ symmetrical metal-based waveguides with a multi-layer triangular graphene-based post. The anisotropic feature for circulation is provided by magnetically-biased graphene in the triangular post. The DC magnetic bias is applied in the z-direction. The magnetized, triangular graphene sheet supports hybrid TM-TE plasmons. To realize the proposed circulator, the structure has been simulated in COMSOL software. In our simulation results, isolation of -40 dB with a transmission loss of -3.5 dB is obtained at the central frequency of 5 THz for a specific design. The bandwidth of the simulated circulator is reported 7.25% with respect to the isolation level of -15 dB. It has been shown that the scattering parameters of the proposed circulator can be changed by altering the chemical potential of the graphene and DC magnetic bias, which makes this circulator a tunable device to be utilized in various plasmonic systems.

## 1. Introduction

In recent years, graphene has opened a new trend in plasmonics due to its remarkable properties [1-3]. One of the important features of graphene is its conductivity, which can be tuned by chemical doping (or electric bias) and external magnetic bias. Based on this feature, a large number of research articles have been reported in the literature that contains various types of THz components such as waveguides [4-14], isolator [15], coupler [16], resonator [17], antennas [18-20], filter [21], Radar Cross-Section (RCS) reduction-based devices [22-24], and graphene-based medical components [25-31]. It should be noted that noble metals support SPPs at the near-infrared and visible frequencies [17, 32, 33]. However, some significant features of graphene-based components, such as extreme confinement, tunable conductivity, and low losses in THz and mid-infrared frequencies differ from any metal-air interface waveguides [34, 35].

Circulators are one of the non-reciprocal devices that are utilized for many applications such as wavelength multiplexers and optical amplifiers [36]. The nonreciprocity of traditional circulators is based on the asymmetrical permeability or permittivity tensors. In the microwave frequencies, ferrites are one of the most important materials for the design of the circulators that circulation occurs due to their permeability tensors [37-42]. In the optical region, some circulators have been introduced in [43-46].

Anisotropic graphene is one of the promising materials that can be utilized for the design of innovative non-reciprocal devices in the THz frequencies. It has a non-symmetrical conductivity tensor when a DC magnetic bias is applied perpendicularly to its surface. This conductivity tensor is tunable via the chemical potential and DC magnetic bias. Graphene-based circulators have been addressed in several research articles [47-51]. In [47], a THz three-port circulator was proposed and investigated. The authors reported a high bandwidth of 29% at the frequency of 17 THz for a low temperature of $T = 3\ K$ [47]. An ultra-wide band graphene-based three-port circulator was introduced in [48] that was designed based on the edge-modes on the graphene. The wide bandwidth of 42% was reported for a DC magnetic bias of 1.5 T and the chemical potential of 0.15 eV in [48]. Another type of three-port graphene-based circulator was suggested in [49], where a circular graphene disk was connected to three graphene-based waveguides. In [50], the authors have utilized coupled-mode theory to theoretically investigate non-reciprocal coupling in a double-layer graphene waveguide on a magneto-optical substrate, which acts as a circulator. A graphene-based junction circulator was studied in [51] where the isolation of 15 dB was obtained for a DC magnetic bias of 0.3 T.

In this paper, we propose and study a novel graphene-based circulator with a multi-layer triangular post. The post is composed of Si-SiO$_2$-Graphene-SiO$_2$-Si layers, where it is located at the junction of three $120^0$ symmetrical metal-based waveguides. To the authors' knowledge, the proposed circulator has not been reported in any published work. The non-reciprocity feature of the device is based on the triangular graphene post that has been magnetically biased in the z-direction. The paper is organized as follows. Section 2 introduces the configuration of the proposed circulator and explains its operation principle. To realize the proposed circulator, it is simulated in the COMSOL software and simulation results are investigated in section 3. It will be shown in this section that the frequency responses of the designed circulator can be tuned by altering the chemical potential of the graphene and the DC magnetic bias. Finally, section 4 concludes the article.

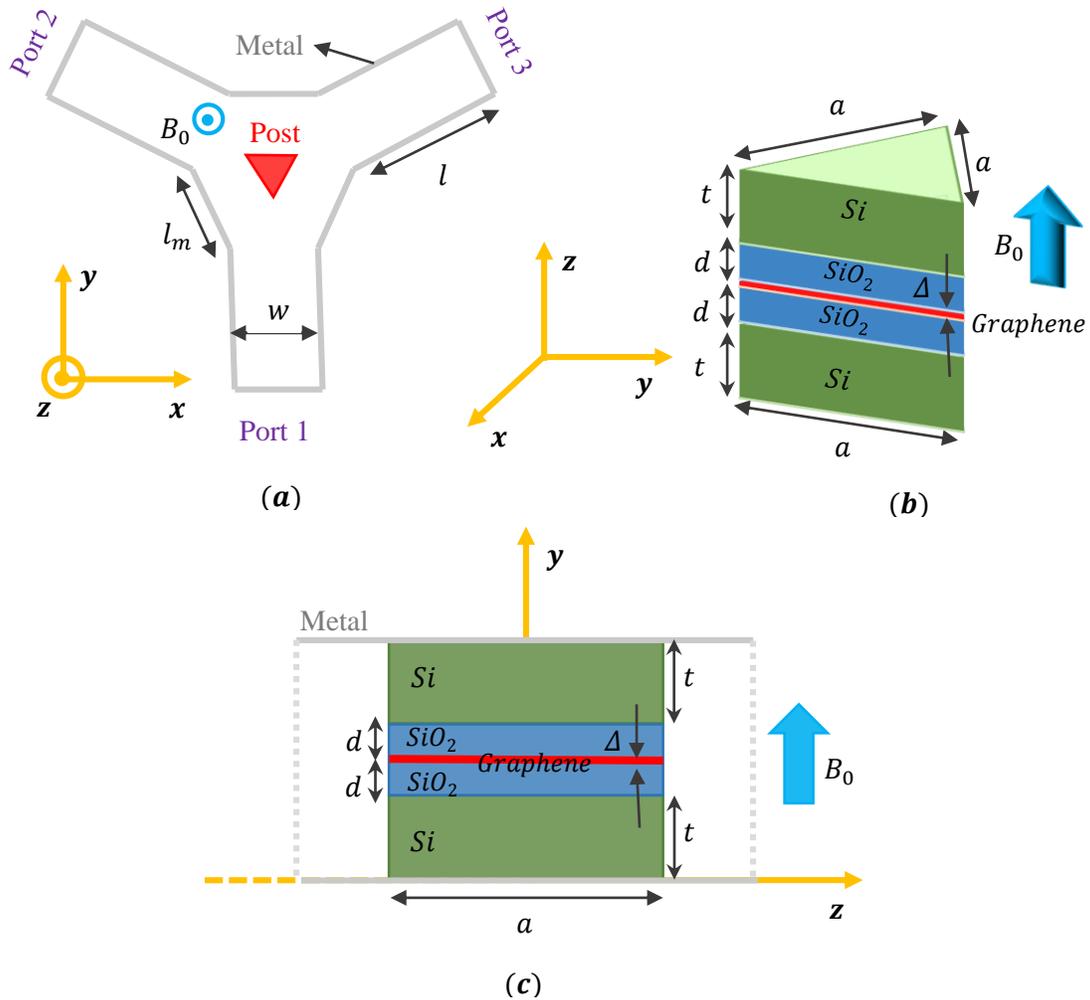

Fig. 1. (a) The top view of the proposed graphene-based circulator, where each port has the dimensions of $w \times (2t + 2d + \Delta)$, (b) The configuration of a triangular post, constituting Si- SiO$_2$-Graphene-SiO$_2$-Si layers, (c) The side view of the proposed circulator.

## 2. The proposed circulator and its operation principle

Fig. 1 illustrates the configuration of the proposed circulator, where three $120^0$ symmetrical metal-based branches have been connected to each other. The graphene-based triangular post is constituted Si-SiO$_2$-Graphene-SiO$_2$-Si layers. A DC magnetic bias is applied in the z-direction. The cross-section of the proposed structure in the x-y and y-z planes have been shown in Fig. 1 (a), (c), respectively.

The operation principle of the proposed device is similar to a microwave three-port circulator with a triangular ferrite post [39-41]. However, there are some differences between them. In the analysis proposed for a microwave circulator with a triangular ferrite post, rigorous field theory has been applied for a full-height ferrite post [39-41]. In this analysis, the microwave circulator with a ferrite post works in TM modes and it is supposed that the post is bounded by perfect magnetic walls (PMC) [39]. The components of TM modes for a ferrite-based circulator are derived by utilizing the familiar method, which is firstly reported in [52]. In our proposed graphene-based circulator, all electromagnetic components of hybrid TM-TE mode exist and thus our structure is a three-dimensional electromagnetic problem. Besides, the circulation of our circulator is based on the asymmetrical conductivity tensor $[\sigma]$ of the magnetized graphene, which has different features compared to the permeability tensor of ferrite in microwave circulators. Therefore, our proposed device cannot be analyzed by the familiar method that exists in the literature for microwave circulators with a triangular ferrite post [39-41, 52]. It should be emphasized that circulation in our structure occurs due to the pairs of resonance modes on the triangular surface of magnetized graphene, which rotation of $30^0$ around the triangular post can be achieved by choosing the appropriate geometrical parameters of the structure and graphene.

In the proposed circulator, the direction of circulation is $1 \to 3 \to 2 \to 1$, described by the following scattering matrix [53]:

$$\bar{\bar{S}} = \begin{pmatrix} S_{11} & S_{12} & S_{13} \\ S_{21} & S_{22} & S_{23} \\ S_{31} & S_{32} & S_{33} \end{pmatrix} \quad (1)$$

Where

$$S_{11} = S_{22} = S_{33} \,,\, S_{32} = S_{21} = S_{13} \,,\, S_{31} = S_{23} = S_{12} \quad (2)$$

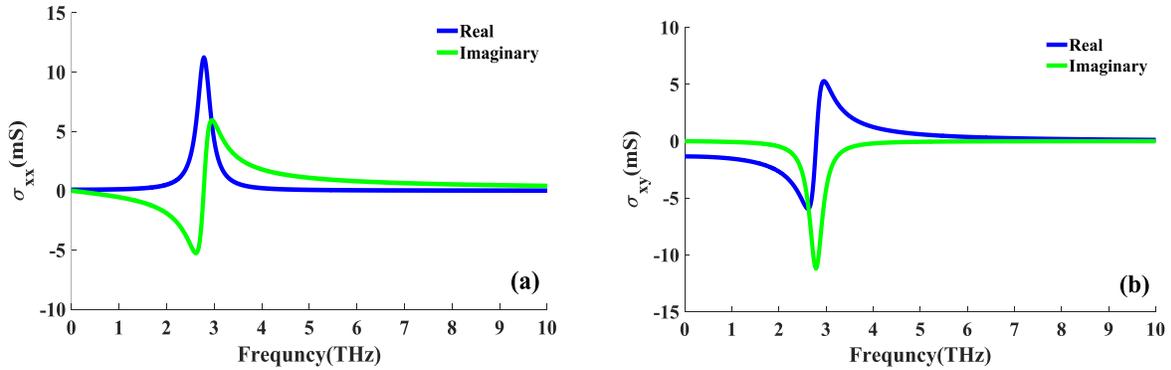

Fig. 2. Real and imaginary parts of: (a) $\sigma_{xx}$, (b) $\sigma_{xy}$ of the graphene conductivity tensor for $B_0 = 3.5\ T, \mu_c = 0.2\ eV, T = 300\ K$.

The conductivity tensor of a magnetically biased graphene sheet in the z-direction can be expressed by a semi-classical model [54]:

$$\bar{\bar{\sigma}} = \begin{pmatrix} \sigma_{xx} & -\sigma_{xy} \\ \sigma_{xy} & \sigma_{xx} \end{pmatrix} \quad (3)$$

Where its diagonal and off-diagonal components are as follows [54]:

$$\sigma_{xx} = \frac{e^2 \mu_c}{\pi \hbar^2} \frac{1/\tau - j\omega}{\omega_c^2 - (\omega + j/\tau)^2} \tag{4}$$

$$\sigma_{xy} = -\frac{e^2 \mu_c}{\pi \hbar^2} \frac{\omega_c}{\omega_c^2 - (\omega + j/\tau)^2} \tag{5}$$

In (5)-(6), the cyclotron frequency is defined:

$$\omega_c = \frac{e B_0 \upsilon_f^2}{\mu_c} \tag{6}$$

In the above relations, $e$ is the electron charge, $\tau$ is the relaxation time, $\hbar$ is the reduced Plank constant, $B_0$ is DC magnetic bias, $\mu_c$ is the chemical potential of the graphene and $\upsilon_f$ is the Fermi velocity ($\upsilon_f \approx 9.5 \times 10^5 \, m/s$) [54].

In Fig. 2, the real and imaginary parts of diagonal and off-diagonal components of the conductivity tensor have been plotted for a graphene sheet with parameters of $B_0 = 3.5 \, T, \mu_c = 0.2 \, eV, T = 300 \, K$. As seen in this figure, the losses (i.e. the parameters of $Re[\sigma_{xx}], Im[\sigma_{xy}]$) occur in the vicinity of the cyclotron frequency. To design the circulator in a frequency band with low losses and high gyrotropic behavior, let us define the gyrotopy and losses:

$$g = \left| \frac{Re[\sigma_{xy}]}{Im[\sigma_{xx}]} \right| \tag{7}$$

$$Losses = |Re[\sigma_{xx}]| + |Im[\sigma_{xy}]| \tag{8}$$

Fig. 3 represents the gyrotropy and losses as a function of frequency for graphene with parameters of $B_0 = 3.5 \, T, \mu_c = 0.2 \, eV, T = 300 \, K$. It can be observed from this figure that $g$ has a maximum value at the frequency of 2.8 THz. However, the losses have large values at this frequency, which means that the circulator should be designed for $f > 2.8 \, THz$ to reduce the propagation losses. To achieve a reasonable level of anisotropy, the gyrotropy should be chosen 0.5-0.6. Therefore, in the frequency range of 4.4-4.8 THz, the losses have negligible values and also the gyrotropy factor is equal to 0.5-0.6.

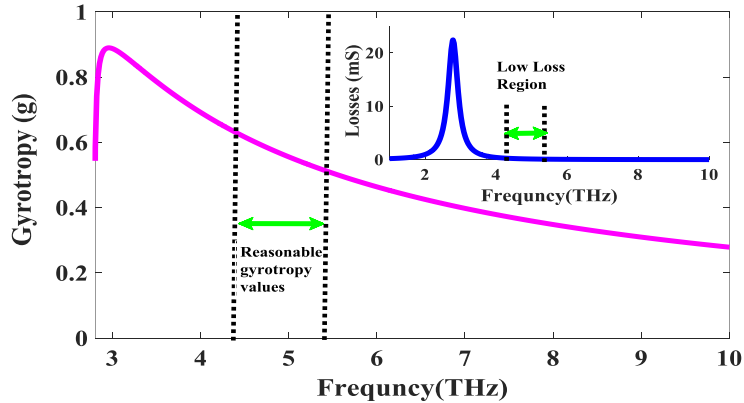

Fig. 3. The gyrotropy and the losses as a function of frequency for the graphene parameters of $B_0 = 3.5 \, T, \mu_c = 0.2 \, eV, T = 300 \, K$.

As explained before, the triangular graphene surface supports a hybrid TM-TE plasmonic mode, where all electromagnetic components exist. Fig. 4 shows the normalized field distributions on the surface of graphene at the frequency of 5 THz for $\mu_c = 0.2 \, eV, T = 300 \, K$. As expected, the degeneracy of two resonant modes (i.e. $f_r^+, f_r^-$) is removed for a magnetized triangular graphene sheet and thus the superposition of these modes ($f_r^+ < f < f_r^-$) can rotate the field to the desired port. By choosing the appropriate geometrical parameters of the structure and the graphene, a rotation of $30^0$ around the triangular post is obtained.

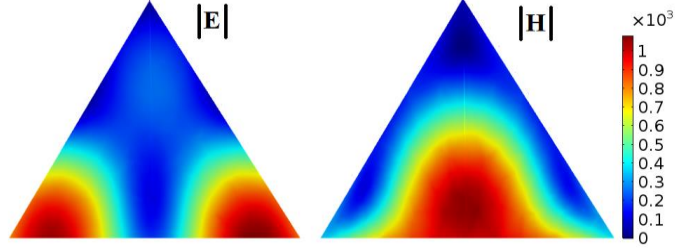

Fig. 4. The magnitude of the electric field vector (left plot) and magnetic field vector (right plot) on the surface of the graphene at the frequency of 5 THz. The graphene parameters are supposed to be $\mu_c = 0.2\ eV, T = 300\ K$.

## 3. Results and Discussion

This section investigates the simulation results of the proposed circulator. In all results, the temperature and the relaxation time are supposed to be $T = 300\ K, \tau = 0.95\ ps$, respectively. The chemical potential of the graphene and the DC magnetic bias are $\mu_c = 0.2\ eV, B_0 = 3.5\ T$, respectively, unless otherwise stated. In our simulation, graphene is considered as a bulk material with the conductivity of $\bar{\bar{\sigma}}_v = \bar{\bar{\sigma}}/\Delta$, where the thickness of graphene is assumed to be $\Delta = 1\ nm$. The relative permittivity of SiO$_2$ and Si layers are $\varepsilon_{SiO_2} = 2.09, \varepsilon_{Si} = 11.9$, respectively. We have utilized copper for three $120^0$ symmetrical metal-based waveguides with the conductivity of $\sigma_{Copper} = 5.8 \times 10^7\ S/m$. The geometrical parameters in our simulation results are $w = 500\ nm, l_m = 450\ nm, l = 1300\ nm, t = 500\ nm, d = 250\ nm$. The side length of the triangle has a value of $a = 200\ nm$ unless otherwise stated.

Fig. 5 demonstrates the scattering parameters of the designed circulator. It is evident from this figure that the transmission loss and isolation are -3.5 dB and -40 dB, respectively. The central frequency of the circulator is 5 THz. The bandwidth of the circulator is about 7.25 % (0.362 THz). Due to the symmetric configuration of our proposed device, the scattering parameters of the circulator excited from the other ports are the same. The magnitude and Z-component of the electric field vector have been shown in Fig. 6 for the simulated circulator. One can see from this figure that the designed circulator works properly as it is excited from port 1.

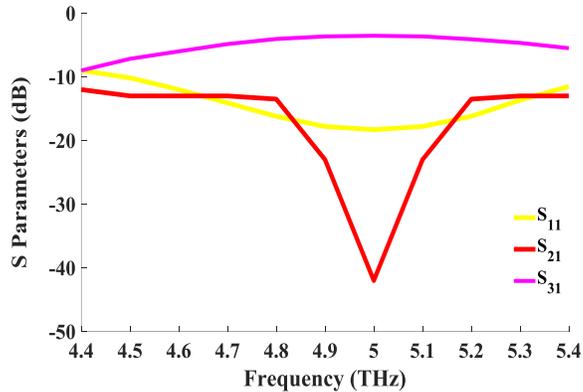

Fig. 5. Scattering parameters of the proposed circulator for the graphene parameters of $B_0 = 3.5\ T, \mu_c = 0.2\ eV, T = 300\ K$.

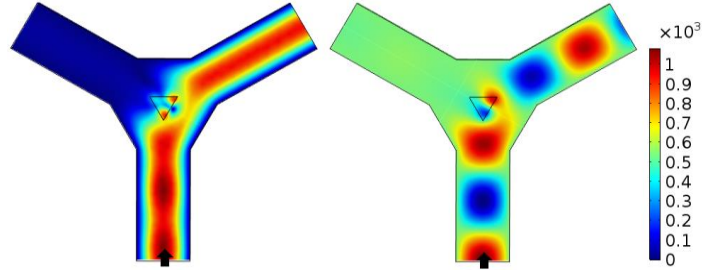

Fig. 6. The magnitude (left plot) and Z- component of the electric field vector (right plot) of the proposed circulator at the frequency of 5 THz. The graphene parameters are assumed to be $B_0 = 3.5\,T, \mu_c = 0.2\,eV, T = 300\,K$.

The designed circulator is a tunable device with adjustable scattering parameters ($|S_{21}|, |S_{31}|$) via the chemical potential and DC magnetic bias. In Fig. 7, the frequency responses of the circulator have been shown for various values of DC magnetic bias. As seen in this figure, the central frequency of the circulator shifts to higher frequencies as the external magnetic bias increases. Furthermore, the isolation loss changes slightly with the increment of the magnetic bias. Fig. 8 represents the scattering parameters of the simulated circulator for various values of chemical potential. By increasing the chemical potential of graphene, the central frequency shifts to lower frequencies, as seen in Fig. 8.

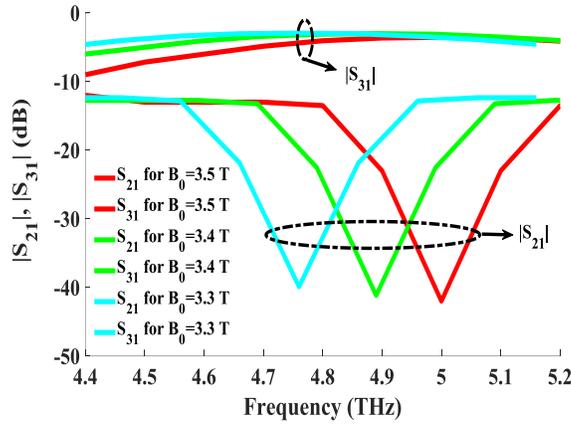

Fig. 7. Scattering parameters of the designed circulator for various values of the DC magnetic bias (B$_0$) for $\mu_c = 0.2\,eV, T = 300\,K$.

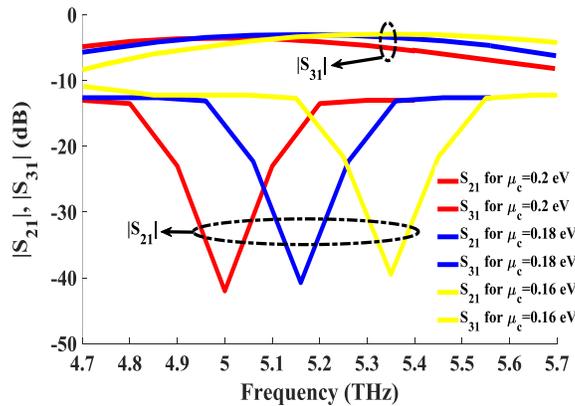

Fig. 8. Scattering parameters of the designed circulator for various values of the chemical potential for $B_0 = 3.5\,T, T = 300\,K$.

One of the important parameters in the design procedure of our proposed circulator is the side length of the triangular post ($a$). Fig. 9 shows the central frequency of the circulator as a function of the side length of the triangle ($a$). It can be observed from this figure that the central frequency has a red-shift for higher values of the side length. Moreover, it should be mentioned that the side length of the triangular post can change the bandwidth of the designed circulator. As seen in Fig. 10, the bandwidth increases slightly as the side length increases. For the side length of $a = 200\ nm$, a bandwidth of 7.25% is achievable. As a final point, it should be noted that the main step in fabricating the proposed circulator is related to the triangular post because the fabrication of three $120^0$ symmetrical metal-based waveguides is similar to ferrite-based circulators. First, a graphene sheet should be deposited on SiO$_2$-Si layers, and then other Si-SiO$_2$ layers are deposited on the obtained Graphene-SiO$_2$-Si layers. Between the Si layers and top and bottom metals, two electromagnets can be placed to magnetically bias the graphene sheet. Electromagnets for generating high magnetic fields have been utilized in many applications such as augmented railguns [55, 56]. Furthermore, a voltage can be applied between the graphene and bottom metal to electrically bias the graphene sheet.

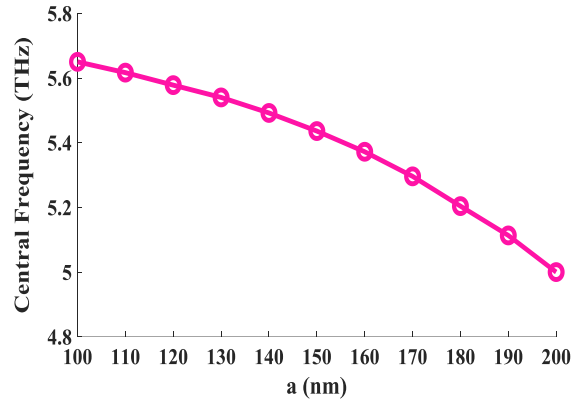

Fig. 9. The central frequency of the proposed graphene-based circulator as a function of the side length ($a$) for the graphene parameters of $B_0 = 3.5\ T, \mu_c = 0.2\ eV, T = 300\ K$.

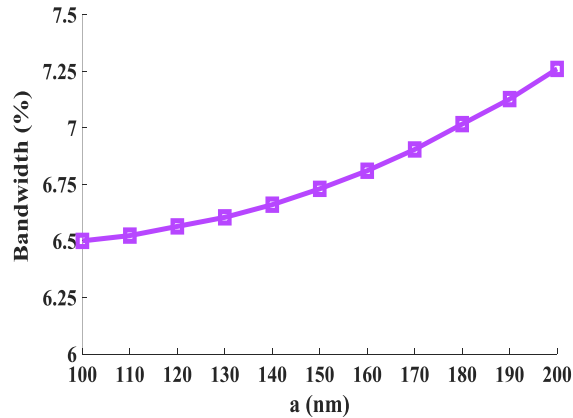

Fig. 10. The Bandwidth of the proposed graphene-based circulator as a function of the side length ($a$) for the graphene parameters of $B_0 = 3.5\ T, \mu_c = 0.2\ eV, T = 300\ K$.

## 4. Conclusion

A three-port THz circulator with a triangular graphene-based post is introduced and investigated in this article. The triangular post is composed of Si-SiO$_2$-Graphene-SiO$_2$-Si layers. The anisotropy effect required for circulation is obtained by a magnetically biased graphene sheet, placed in the triangular post. Our designed circulator can be integrated with other traditional THz devices due to three metallic branches. It has been shown that the central frequency of the simulated circulator can be tuned by altering the chemical potential and external magnetic bias. In a specific design at the central frequency of 5 THz, the bandwidth of 7.25 % with the isolation of -40 dB and the transmission loss of -3.5 dB is achieved. Altering the side length of the triangular post can enhance the bandwidth of the device. The presented study can be utilized for the design of innovative graphene-based circulators in the THz frequencies.